**Magnetic oxide semiconductors**

T Fukumura, H Toyosaki and Y Yamada

Institute for Materials Research, Tohoku University, Sendai 980-8577, Japan

Magnetic oxide semiconductors, oxide semiconductors doped with transition metal elements, are one of the candidates for a high Curie temperature ferromagnetic semiconductor that is important to realize semiconductor spintronics at room temperature. We review in this paper recent progress of researches on various magnetic oxide semiconductors. The magnetization, magneto-optical effect, and magneto-transport such as anomalous Hall effect are examined from viewpoint of feasibility to evaluate the ferromagnetism. The ferromagnetism of Co-doped $TiO_2$ and transition metal-doped ZnO is discussed.

E-mail: fukumura@imr.tohoku.ac.jp

1. Introduction

Magnetic semiconductor is the semiconductor doped with spin as a new degree of freedom of electron, and is expected to be a promising candidate material for next generation spintronics devices utilizing electronically or optically controlled magnetism. II-VI semiconductor based magnetic semiconductors have been studied for over two decades [1]. The Mn ion has been often used as a spin injector, yielding giant magnetoresistance and magneto-optical effect, the latter of which is used for the practical application of (Cd,Mn)Te to an optical isolator [2]. Several tens of molar percents of Mn can be doped into II-VI semiconductors, whereas the electron density is only $10^{19}$ cm$^{-3}$ at most and the hole doping is difficult. These drawbacks result in the quite low ferromagnetic Curie temperature ($T_C$) in the order of ~1 K [3], that is well explained by the carrier induced mechanism [4]. On the other hand, a III-V semiconductor, GaAs doped with only 5 mol% of Mn, has the $T_C$ over 100 K due to the strong $p$-$d$ exchange interaction intermediated by the mobile holes [5]. Furthermore, the light induced ferromagnetism [6], the injection of polarized spins into semiconductor [7,8], and the modulation of $T_C$ by electric field effect [9] have been demonstrated. These features are a consequence of the controllable carrier density for the ferromagnetic semiconductor in contrast with the other well-known ferromagnetic conductors as listed in Table 1. For the practical use of devices at room temperature, however, the ferromagnetic $T_C$ beyond room temperature is needed.

Oxide semiconductors have been extensively used as transparent electrodes so far [10]. Various kinds of oxide semiconductors are listed in Table 2. They have several interesting features from viewpoint of industrial applications as seen in Table 3. Recent progress of growth techniques for oxide thin films enables us to treat the oxide semiconductors as conventional compound semiconductors. Therefore, oxide semiconductors can be host compounds for magnetic semiconductors. Indeed, it was pointed out that the capability of high electron doping and the rather heavy effective electron mass for the oxide semiconductors could be quite efficient to realize high $T_C$ [11]. Consequently, room temperature ferromagnetic semiconductors, anatase and rutile phases Co-doped $TiO_2$, were discovered with combinatorial approach [12,13]. Many studies have emerged and reported on the discovery of high $T_C$ ferromagnetic oxide semiconductors so far.

This paper reviews the recent reports on the magnetic oxide semiconductors although some of them might be inevitably omitted due to the rapidly increasing number of the reports. There have been several reviews for these topics elsewhere [14-18]. Various ferromagnetic oxide semiconductors have been reported so far, however, the ferromagnetism has been still of strong debate. In order to examine the ferromagnetism carefully, several considerations on the evaluation from experimental side are given. Especially, the magnetism for transition metal-doped ZnO and Co-doped $TiO_2$ is discussed.

2. Various oxide semiconductors

2.1. ZnO

Mn-doped ZnO was reported as a new class of II-VI magnetic semiconductor [11]. The properties were similar to the typical II-VI magnetic semiconductors [11,19]: the absorption due to *d-d* transition of Mn ion, the large magnetoresistance at low temperature, and the spin glass magnetic behavior. Figure 1 shows a diagram summarizing the existing II-VI magnetic semiconductors. II-VI magnetic oxide semiconductors are added to the conventional ones, although Mn-doped CdO and HgO have not been reported. ZnO doped with the other transition metals were synthesized with combinatorial approach [20,21]. (It is noted that various transition metals can be easily doped.) They were not ferromagnetic [21] and Mn-, Co-, Ni-, and Cu-doped ZnO showed the appreciable magnetic circular dichroism (MCD) at the absorption edge at low temperature [22]. Among them, Co-doped ZnO showed the largest MCD, that was comparable with that of an optical isolator material, Mn-doped CdTe [23]. Photoemission spectroscopy revealed that Mn-doped ZnO had much larger *p-d* exchange constant [24], and X-ray absorption spectroscopy also confirmed the large *p-d* exchange constant for Mn-, Fe-, and Co-doped ZnO [25]. These results are consistent with their large MCD. Hysteretic magnetoresistance for Mn-doped ZnO appeared at ultra-low temperature, implying emergence of the long range magnetic ordering at ultra-low temperature [26,27].

There have been many reports on the fabrication of transition metal-doped ZnO. Both bulk and thin film specimens have been synthesized. So far, Ti-doped [20,21,28], V-doped [20,21,29,30], Cr-doped [20,21,31,32], Mn-doped [11,19-21,32-57], Fe-doped [20,21,54,55,58], Co-doped [20,21,30,32,52-55,59-82], Ni-doped [20,21,32,56,83-85, Cu-doped [20,21,59,86], (Mn,Sn)-doped [87], (Fe,Co)-doped [88], (Fe,Cu)-doped [58,89], and (Li,Co)-doped [90] ZnO have been reported. They were reported to be both ferromagnetic and nonferromagnetic even for the same compound as described in Section 3.

2.2. $TiO_2$

Co-doped anatase and rutile $TiO_2$, that have high transparency under visible light,

were reported to be ferromagnetic above 400 K [12,13]. Co-doped anatase $TiO_2$ showed the degenerate semiconducting behavior [12] and the MCD was very large and ferromagnetic at room temperature [91]. For Co-doped rutile $TiO_2$, the anomalous Hall effect was observed at room temperature [92-94] and the Hall conductivity was found to be a systematic function of the electron density [95]. Many studies on Co-doped $TiO_2$ have been reported so far [96-119]. X-ray absorption spectroscopy, X-ray photoemission spectroscopy, extended X-ray absorption fine structure, X-ray absorption near the edge structure, and X-ray MCD have been performed to determine the valence of Co ion in $TiO_2$ [120-127]. However, the present results have not been conclusive to determine whether $Co^{2+}$ ion substituted for Ti site or Co metal precipitated.

The other transition metals have also been doped. Cr-doped rutile $TiO_2$ was reported to be ferromagnetic above 400 K [128]. V-doped anatase $TiO_2$ was reported to be ferromagnetic at room temperature [129]. Fe-doped rutile $TiO_2$ was reported to be ferromagnetic over room temperature [130-133] and to show anomalous Hall effect at room temperature [131,132]. Fe-doped anatase $TiO_2$ was reported to be ferromagnetic at room temperature [134,135]. Ni-doped anatase $TiO_2$ was also reported to be ferromagnetic at room temperature [135]. The transition metal-doped $TiO_2$ has been often reported to be ferromagnetic, however, several papers pointed out the possible extrinsic origin of the ferromagnetism as described in Section 3.

2.3. $SnO_2$

Mn-doped $SnO_2$ was reported to show large magnetoresistance at low temperature and paramagnetic behavior [136-138]. On the other hand, the same Mn-doped $SnO_2$ was reported to be ferromagnetic with $T_C$ = 340 K [139]. Fe-doped $SnO_2$ was reported to be ferromagnetic with $T_C$ = 360 K [139] and $T_C$ = 610 K [140]. Co-doped $SnO_2$ was reported to be ferromagnetic with $T_C$ = 650 K and rather high magnetization with 7.5$\mu_B$ [141].

2.4. $In_2O_3$

Mn- and Fe-doped $In_2O_3$ were reported to be ferromagnetic at room temperature [142]. Mn-doped indium-tin oxide was reported to show anomalous Hall effect at room temperature [143].

3. Examination of ferromagnetism

3.1. Brief summary of transition metal-doped ZnO and Co-doped $TiO_2$

ZnO and $TiO_2$ have been most extensively studied. ZnO has been doped with various transition metals, whereas $TiO_2$ has been doped mostly with Co. Here we summarize the magnetism of the transition metal-doped ZnO and the Co-doped $TiO_2$ that has been reported so far.

ZnO doped with transition metals were reported to be nonferromagnetic for the first time [19,21]. Subsequently, ZnO doped with various transition metals was often reported to be ferromagnetic as summarized in Table 4. The range of the $T_C$ is very broad: from 2 K to over room temperature. About 40 % of the total papers have claimed the ferromagnetism. Among 3$d$ transition metals, Co-doped and Mn-doped ZnO have been frequently reported to be ferromagnetic. On the whole, $T_C$ of Co-doped ZnO is higher than that of Mn-doped ZnO.

For Co-doped $TiO_2$, the high temperature ferromagnetism was reported for the first time [12,13]. Subsequent studies often reported the high $T_C$ also. The $T_C$ lower than room temperature has scarcely been reported in contrast with ZnO based magnetic semiconductors. However, several groups claimed the extrinsic mechanism for the ferromagnetism such as Co metal precipitations as listed in Table 5. Most of them reported the appearance of a cluster structure in $TiO_2$ matrix by using transmission electron microscope.

3.2. Remarks on fabrication methods

There have been various techniques to fabricate magnetic oxide semiconductors [18]. Thermally equilibrium solubility of transition metal in ZnO is relatively high so that the fabrication of the bulk specimen is sometimes possible, although the fabrication of the thin film specimen has been often employed. On the other hand, the solubility of Co in $TiO_2$ is very low. Accordingly, thermally nonequilibrium processes such as thin film growth process have been usually performed.

High temperature growth sometimes leads to high crystalline quality of thin film, however, such condition may be close to thermally equilibrium state, where the solubility of dopant is limited. On the other hand, low temperature processing sometimes led to appearance of metastable ferromagnetic secondary phases [40,57]. In addition to the growth process, heat treatment subsequent to the growth has to be noticed. Post annealing temperature similar with or higher than the growth temperature may yield segregation of secondary phases.

As for the structural properties, some of them have reported on the appearance of impurity phase or precipitation, but the others have not. If any precipitation, it is necessary to investigate the origin of the magnetism. Therefore, the examination of the various magnetic properties would be more straightforward to determine the magnetism as described below.

## 3.3. Magnetization

Magnetization measurements have been often used as a proof of the ferromagnetism because the magnetization is one of the most fundamental properties of ferromagnets. However, it is difficult to distinguish the intrinsic ferromagnetism from the extrinsic one such as ferromagnetic precipitations or secondary phases (Fig. 2(a)), because all the magnetization signals are measured together. Accordingly, only magnetization measurement is not enough to prove the intrinsic ferromagnetism. Magnetic resonances as well as various spectroscopies can probe the chemical state, thus can rule out the possible metal precipitation. However, they may not separate the magnetic signal of the bulk from that of nonmetal precipitation having similar chemical states such as oxide.

## 3.4. Magneto-optical effect

Generally, in magnetic semiconductors, the *s*- and *p*- charge carriers in the host semiconductor couple with the *d*-electrons of the transition metal ions via the exchange interaction. Consequently, photoexcited charge carriers in the host semiconductor show the sizable magneto-optical effect: the magneto-optical spectrum reflects the energy band structure of the host semiconductor, where a large magneto-optical effect emerges around the absorption edge (Fig. 2(b) right). MCD is the difference of the absorption coefficients for two circularly polarized lights and is proportional to the energy derivative of the absorption coefficient [144] so that the MCD is the largest around the onset of the absorption. Accordingly, the MCD spectrum can identify the ferromagnetic origin because the ferromagnetic metal often shows a monotonous MCD spectrum [91].

Figure 3(a) shows the MCD and absorption spectra for Co-doped ZnO at 5 K (dashed lines) and Co-doped anatase TiO$_2$ at room temperature (solid lines) [22,23,91]. ZnO is a direct transition semiconductor, thus the relationship between the MCD and absorption spectra is apparent: the sharp and large MCD signal appears at the abrupt increase in the absorption [145]. The MCD signal around 2 eV corresponds to the *d-d* level absorption of Co$^{2+}$ ion [21], that is hardly seen in the scale of Fig. 3(a). The magnetic field dependence of the MCD is linear at any photon energy as shown in Fig. 3(b), hence the magnetism is paramagnetic.

Co-doped anatase TiO$_2$ shows a little broad and large MCD signal around the absorption edge and below the absorption edge. The magnetic field dependence of the MCD shows ferromagnetic hysteresis loop as shown in Fig. 3(b), where the dependence at any photon energy is identical (except for the magnitude) [91]. In addition, there is not any other superposed magnetic signal such as paramagnetic signal. Therefore, the MCD is originated from single ferromagnetic source. In other word, if the ambiguous relationship between the MCD and absorption spectra, the different magnetic field dependence depending on photon energy, or the superposition of several magnetic behaviors can be seen, the ferromagnetism might be extrinsic [146].

3.5. Anomalous Hall effect

Anomalous Hall effect is a ferromagnetic response of charge carriers in electrically conductive ferromagnets: emergence of Hall voltage proportional to the magnetization, attributed to asymmetric carrier scattering by magnetic impurities in the presence of spin-orbit interaction (Fig. 2(c) right). Hall resistivity ($\rho_{xy}$) in ferromagnets is generally expressed as $\rho_{xy} = R_o B + R_s \mu_0 M$ ($R_o$: ordinary Hall coefficient, *B*: magnetic induction, $R_s$: anomalous Hall coefficient, $\mu_0$: magnetic permeability, $M$: magnetization) [147]. The first term denotes ordinary Hall effect that is proportional to inverse of the carrier density, and the second term denotes anomalous Hall effect that is proportional to the magnetization.

Anomalous Hall effect represents ferromagnetic spin polarization of charge carrier. Thus, the observation of anomalous Hall effect was recognized as a strong evidence of the intrinsic ferromagnetism for the ferromagnetic semiconductors. However, Shinde et al. [94] pointed out that anomalous Hall effect could be observed for nonmagnetic material embedded with magnetic clusters, the density of which was around the bulk percolation threshold [148]. The nonmagnetic material embedded with magnetic clusters may show an appreciable magnetoresistance in accordance with the saturation of the magnetization [149], implying the spin dependent scattering of the charge carriers between the magnetic clusters (Fig. 2(c) center). Co-doped rutile $TiO_2$ shows anomalous Hall effect, the magnetic field dependence of which coincides with that of magnetization as shown in Fig. 4. On the other hand, the magnetoresistance gradually changes with increasing magnetic field, irrespective of the saturation of the magnetization [95]. The Co content is 3 mol%, thus is sufficiently below the percolation limit. Therefore, it is difficult to assume the spin conservative conduction between the spatially isolated ferromagnetic clusters as is observed in the granular materials.

Figure 5 shows the mapping of the appearance of ferromagnetic Co-doped rutile $TiO_2$ as functions of the conductivity and Co content, determined from the anomalous Hall effect and MCD measurements [95,150]. For the higher conductivity and Co content, the ferromagnetic MCD and AHE appear simultaneously. This result represents that the ferromagnetism is triggered by the increase of the carrier density and Co content and is originated from the single ferromagnetic source.

Figure 6 shows the relationship between the anomalous Hall conductivity and conductivity (i.e. proportional to the carrier density) for the ferromagnetic Co-doped rutile $TiO_2$, where the data for the different Co contents, carrier densities, and measurement temperatures are plotted [95]. This relationship is approximately scaled for ten-thousand-fold range of the conductivity as $\sigma_{AHE} \propto \sigma_{xx}^{1.5-1.7}$, that is different from the skew scattering or side jump mechanisms, $\propto \sigma_{xx}$ or $\propto \sigma_{xx}^2$, respectively. In Fig. 6, the two data obtained from Ref. [93] are also plotted. The plots almost overlap the scaling, implying that the conduction mechanism is the same as Ref. [95] and that the precipitation observed did not take significant role for the magneto-transport (Fig. 2(c)).

4. Problems to be solved

For III-V ferromagnetic semiconductors, there have been appreciable achievements using heterostructure devices so far. On the other hand, the ferromagnetic oxide semiconductors are still in an early stage. In order to make further progress, the following problems are to be solved.

For Co-doped $TiO_2$, the ferromagnetic MCD has been observed for anatase and rutile phases [91,150]. The coincidence between the broad MCD and absorption spectra is not perfect in comparison with Co-doped ZnO in Fig. 3. Some theoretical approaches to justify the MCD spectra are needed in spite of the complicated energy band structure. The MCD disappeared for both phases with decreasing the carrier density. The appearance of AHE synchronizes with that of MCD for rutile phase, but AHE for anatase phase has not been reported so far. The appearance of AHE for anatase phase, if any, has to be confirmed. The carrier density for rutile phase is larger than that for anatase phase, hence higher carrier doping may be necessary for anatase phase. Also, the mechanism of AHE has to be clarified in order to understand the difference from that for the granular systems.

For transition metal-doped ZnO, the dependence of the ferromagnetism on the carrier density has not been reported although the *n*-type carrier doping can be easily done by chemical doping. The ferromagnetism has to be controlled by the carrier density. Furthermore, the ferromagnetism has to be examined from independent systematic measurements such as magnetization and MCD. It is noted that ultralow temperature magnetoresistance for Mn-doped ZnO showed a hysteresis, suggesting the long range magnetic ordering, without appearance of anomalous Hall effect [26,27], probably due to the small spin-orbit coupling in ZnO.

5. Conclusion

Recent progress of various magnetic oxide semiconductors is briefly reviewed. There has been significant controversy regarding not only the validity of ferromagnetism but also the origin. This issue has to be settled thoroughly by further studies. Nevertheless, it is noted that anomalous Hall effect controlled by the electron density has appeared for Co-doped rutile $TiO_2$ [95], because it would directly lead to the electrical control. Operation of spintronics devices made of magnetic oxide semiconductor, hopefully at higher temperature, is promising.

**References**

[1]     Giriat W and Furdyna J K 1988 *Semiconductors and Semimetals* vol 25 ed Willardson R K and Beer A C (Boston: Academic Press) p 1

[2]     Onodera K, Masumoto T and Kimura M 1994 *Electron. Lett.* **30** 1954

[3]     Haury A, Wasiela A, Arnoult A, Cibert J, Tatarenko S, Dietl T and Merle d'Aubigné Y 1997 *Phys. Rev. Lett.* **79** 511

[4]     Dietl T, Haury A and Merle d'Aubigne Y 1997 *Phys. Rev.* B **55** R3347

[5]     Ohno H 1998 *Science* **281** 951

[6]     Koshihara S, Oiwa A, Hirasawa M, Katsumoto S, Iye Y, Urano C, Takagi H and Munekata H 1997 *Phys. Rev. Lett.* **78** 4617

[7]     Fiederling R, Keim M, Reuscher G, Ossau W, Schmidt G, Waag A and Molenkamp L W 1999 *Nature* **402** 787

[8]     Ohno Y, Young D K, Beschoten B, Matsukara F, Ohno H and Awschalom D D 1999 *Nature* **402** 790

[9]     Ohno H, Chiba D, Matsukura F, Omiya T, Abe E, Dietl T, Ohno Y and Ohtani K 2000 *Nature* **408** 944

[10]    Ueda K and Hosono H *Toumei Doudenmaku no Gijutsu* ed Nihon Gakujutsu Shinkoukai (Tokyo: Ohm-sya) [in Japanese]

[11]    Fukumura T, Jin Z, Ohtomo A, Koinuma H and Kawasaki M 1999 *Appl. Phys. Lett.* **75** 3366

[12]    Matsumoto Y, Murakami M, Shono T, Hasegawa T, Fukumura T, Kawasaki M, Ahmet P, Chikyow T, Koshihara S and Koinuma H 2001 *Science* **291** 854

[13]    Matsumoto Y, Takahashi R, Murakami M, Koida T, Fan X J, Hasegawa T,

Fukumura T, Kawasaki M, Koshihara S and Koinuma H 2001 *Jpn. J. Appl. Phys.* **40** L1204

[14] Fukumura T, Kawasaki M, Jin Z, Kimura H, Yamada Y, Haemori M, Matsumoto Y, Inaba K, Murakami M, Takahashi R, Hasegawa T and Koinuma H 2002 *MRS Proc.* **700** 45

[15] Chambers S A and Farrow R F C 2003 *MRS Bulletin* **28** 729

[16] Matsumoto Y, Koinuma H, Hasegawa T, Takeuchi I, Tsui F and Yoo Y K 2003 *MRS Bulletin* **28** 734

[17] Prellier W, Fouchet A and Mercey B 2003 *J. Phys.: Condens. Matter* **15** R1583

[18] Fukumura T, Yamada Y, Toyosaki H, Hasegawa T, Koinuma H and Kawasaki M 2004 *Appl. Surf. Sci.* **223** 62

[19] Fukumura T, Jin Z, Kawasaki M, Shono T, Hasegawa T, Koshihara S and Koinuma H 2001, *Appl. Phys. Lett.* **78**, 958

[20] Jin Z, Murakami M, Fukumura T, Matsumoto Y, Ohtomo A, Kawasaki M and Koinuma H 2000 *J. Cryst. Growth* **214/215** 55

[21] Jin Z, Fukumura T, Kawasaki M, Ando K, Saito H, Sekiguchi T, Yoo Y Z, Murakami M, Matsumoto Y, Hasegawa T and Koinuma H 2001 *Appl. Phys. Lett.* **78** 3824

[22] Ando K, Saito H, Jin Z, Fukumura T, Kawasaki M, Matsumoto Y and Koinuma H 2001 *J. Appl. Phys.* **89** 7284

[23] Ando K, Saito H, Jin Z, Fukumura T, Kawasaki M, Matsumoto Y and Koinuma H 2001 *Appl. Phys. Lett.* **78** 2700

[24] Mizokawa T, Nambu T, Fujimori A, Fukumura T and Kawasaki M 2002 *Phys. Rev. B* **65** 085209

[25] Okabayashi J, Ono K, Mizuguchi M, Oshima M, Gupta S S, Sarma D D, Mizokawa T, Fujimori A, Yuri M, Chen C T, Fukumura T, Kawasaki M and Koinuma H 2004 *J. Appl. Phys.* **95** 3573

[26] Andrearczyk T, Jaroszynski J, Sawicki M, van Khoi L, Dietl T, Ferrand D, Bourgognon C, Cibert J, Tatarenko S and Fukumura T 2001 *Springer Proc. Phys.* **87** Part 1 234

[27] Sawicki M, van Khoi L, Matsukura F, Dietl T, Fukumura T, Jin Z, Koinuma H and Kawasaki M 2003 *J. Supercond.* **16** 147

[28] Park Y R and Kim K J 2003 *Solid State Commun.* **123** 147

[29] Saeki H, Tabata H and Kawai T 2001 *Solid State Commun.* **120** 439

[30] Tabata H, Saeki M, Guo S L, Choi J H and Kawai T 2001 *Physica* B **308** 993

[31] Satoh I and Kobayashi T 2003 *Appl. Surf. Sci.* **216** 603

[32] Ueda K, Tabata H and Kawai T 2001 *Appl. Phys. Lett* **79** 988

[33] Jung S W, An SJ, Yi GC, Jung C U, Lee S I and Cho S 2002 *Appl. Phys. Lett.* **80** 4561

[34] Kim J H, Lee J B, Kim H, Choo W K, Ihm Y and Kim D 2002 *Ferroelectrics* **273** 71

[35] Kolesnik S, Dabrowski B and Mais J 2002 *J. Supercond.* **15** 251

[36] Edahiro T, Fujimura N and Ito T 2003 *J. Appl. Phys.* **93** 7673

[37] Cheng X M and Chien C L 2003 *J. Appl. Phys.* **93** 7876

[38] Kim K J and Park Y R 2003 *J. Appl. Phys.* **94** 867

[39] Zhou H J, Hofmann D M, Hofstaetter A and Meyer B K 2003 *J. Appl. Phys.* **94** 1965

[40] Han S J, Jang T H, Kim Y B, Park B G, Park J H and Jeong Y H 2003 *Appl. Phys.*

*Lett.* **83** 920

[41]  Sharma P, Gupta A, Rao K V, Owens F J, Sharma R, Ahuja R, Guillen J M O, Johansson B and Gehring G A 2003 *Nat. Mater.* **2** 673

[42]  Chang Y Q, Wang D B, Luo X H, Xu X Y, Chen X H, Li L, Chen C P, Wang R M, Xu J and Yu D P 2003 *Appl. Phys. Lett.* **83** 4020

[43]  Chang Y Q, Luo X H, Xu X Y, Li L, Chen J P, Wang R M and Yu D P 2003 *Chin. Phys. Lett.* **20**

[44]  Kim D S, Lee S, Min C, Kim H M, Yuldashev S U, Kang T W, Kim D Y and Kim T W 2003, *Jpn. J. Appl. Phys.* **42**, 7217

[45]  Kim S S, Moon J H, Lee B T, Song O S and Je J H 2004 *J. Appl. Phys.* **95** 454

[46]  Kim Y M, Yoon M, Park I W, Park Y J and Lyou J H 2004 *Solid State Commun.* **129** 175

[47]  Roy V A L, Djurisic A B, Liu H, Zhang X X, Leung Y H, Xie M H, Gao J, Lui H F and Surya C 2004 *Appl. Phys. Lett.* **84** 756

[48]  Lim S W, Jeong M C, Ham M H, Myoung J M 2004 *Jpn. J. Appl. Phys.* **43** L280

[49]  Heo Y W, Ivill M P, Ip K, Norton D P, Pearton S J, Kelly J G, Rairigh R, Hebard A F and Steiner T 2004 *Appl. Phys. Lett.* **84** 2292

[50]  Viswanatha R, Sapra S, Gupta S S, Satpati B, Satyam P V, Dev B N and Sarma D D 2004 *J. Phys. Chem.* B **108** 6303

[51]  Savchuk A I, Gorley P N, Khomyak V V, Ulyanytsky K S, Bilichuk S V, Perrone A and Nikitin P I 2004 *Mat. Sci Eng.* B **109** 196

[52]  Theodoropoulou N A, Hebard A F, Norton D P, Budai J D, Boatner L A, Lee J S, Khim Z G, Park Y D, Overberg M E, Pearton S J and Wilson, R G 2003 *Solid State Electron.* **47** 2231

[53] Ip K, Frazier R M, Heo Y W, Norton D P, Abernathy C R, Pearton S J, Kelly J, Rairigh R, Hebard A F, Zavada J M and Wilson R G 2003 *J. Vac. Sci. Tech.* B **21** 1476

[54] Yoon S W, Cho S B, We S C, Yoon S, Suh B J, Song H K, Shin Y J 2003, *J. Appl. Phys.* **93**, 7879

[55] Kolesnik S, Dabrowski B and Mais J 2004, *J. Appl. Phys.* **95**, 2582

[56] Schwartz D A, Norberg N S, Nguyen Q P, Parker J M and Gamelin D R 2003, *J. Am. Chem. Soc.* **125**, 13205

[57] Kundaliya D C, Ogale S B, Lofland S E, Dhar S, Metting C J, Shinde S R, Ma Z, Varughese B, Ramanujachary K V, Salamanca-Riba L and Venkatesan T 2004, *Nat. Mater.* **3,** 709

[58] Guzman E, Hochmuth H, Lorenz M, von Wenckstern H, Rahm A, Kaidashev E M, Ziese M, Setzer A, Esquinazi P, Poppl A, Spemann D, Pickenhain, R, Schmidt H and Grundmann M 2004 *Annal. Phys.* **13** 57

[59] Brumage W H, Dorman C F and Quade CR 2001 *Phys. Rev.* B **63** 104411

[60] Kim K J and Park Y R 2002 *Appl. Phys. Lett.* **81** 1420

[61] Kim J H, Lee J B, Kim H, Kim D, Ihm Y and Choo W K 2002 *IEEE Trans. Magn.* **38** 2880

[62] Yang S G, Pakhomov A B, Hung S T and Wong C Y S 2002 *IEEE Trans. Magn.* **38** 2877

[63] Kim J H, Kim H, Kim D, Ihm Y E and Choo W K 2002 *J. Appl. Phys.* **92** 6066

[64] Lee H J, Jeong S Y, Cho C R and Park C H 2002 *Appl. Phys. Lett.* **81** 4020

[65] Radovanovic P V, Norberg N S, McNally K E, Gamelin D R 2002 *J. Am. Chem. Soc.* **124** 15192

[66]   Lim S W, Hwang D K, Myoung J M 2003 *Solid State Commun.* **125** 231

[67]   Kim J H, Choo W K, Kim H, Kim D and Ihm Y 2003 *J. Korea. Phys. Soc.* **42** S258

[68]   Kim J H, Kim H, Kim D, Ihm Y E and Choo W K 2003 *Physica* B **327** 304

[69]   Rode K, Anane A, Mattana R, Contour J P, Durand O and LeBourgeois R 2003 *J. Appl. Phys.* **93** 7676

[70]   Pakhomov A B, Roberts B K and Krishnan K M 2003 *Appl. Phys. Lett.* **83** 4357

[71]   Qi Z M, Li A X, Su F L, Zhou S M, Liu Y M, Zhao Z Y 2003 *Mat. Res. Bull* **38** 1791

[72]   Ren M J, Yan S S, Ji G, Chen Y X, Song H Q, Mei L M 2003 *Chin. Phys. Lett.* **20** 2239

[73]   Risbud AS, Spaldin N A, Chen Z Q, Stemmer S, Seshadri R 2003, *Phys. Rev.* B **68**, 205202

[74]   Norton D P, Overberg M E, Pearton S J, Pruessner K, Budai J D, Boatner L A, Chisholm M F, Lee J S, Khim Z G, Park Y D and Wilson R G 2003 *Appl. Phys. Lett.* **83** 5488

[75]   Park J H, Kim M G, Jang H M, Ryu S and Kim Y M 2004 *Appl. Phys. Lett.* **84** 1338

[76]   Jedrecy N, von Bardeleben H J, Zheng Y and Cantin J L 2004 *Phys. Rev.* B **69** 41308

[77]   Kim J H, Kim H, Kim D, Ihm Y and Choo W K 2004 *J. Euro. Ceram. Soc.* **24** 1847

[78]   Yan S S, Ren C, Wang X, Xin Y, Zhou Z X, Mei L M, Ren M J, Chen Y X, Liu Y H and Garmestani H 2004 *Appl. Phys. Lett.* **84** 2376

[79] Wi S C, Kang J S, Kim J H, Cho S B, Kim B J, Yoon S, Suh B J, Han S W, Kim K H, Kim K J, Kim B S, Song H J, Shin H J, Shim J H and Min B I 2004 *Appl. Phys. Lett.* **84** 4233

[80] Fouchet A, Prellier W, Padhan P, Simon C, Mercey B, Kulkarni V N and Venkatesan T 2004 *J. Appl. Phys.* **95** 7187

[81] Pakhomov A B, Roberts B K, Tuan A, Shutthanandan V, McCready D, Thevuthasan S, Chambers S A and Krishnan K M 2004 *J. Appl. Phys.* **95** 7393

[82] Prellier W, Fouchet A, Simon C and Mercey B 2004 *Mat. Sci Eng.* B **109** 192

[83] Wakano T, Fujimura N, Morinaga Y, Abe N, Ashida A and Ito T 2001 *Physica* E **10** 260

[84] Jung S W, Park W I, Yi G C and Kim M 2003 *Adv. Mat.* **15** 1358

[85] Radovanovic P V and Gamelin D R 2003 *Phys. Rev. Lett.* **91** 157202

[86] Leiter F, Zhou H, Henecker F, Hofstaetter A, Hofmann D M and Meyer B K 2001 *Physica* B **308** 908

[87] Norton D P, Pearton S J, Hebard A F, Theodoropoulou N, Boatner L A and Wilson R G 2003 *Appl. Phys. Lett.* **82** 239

[88] Cho Y M, Choo W K, Kim H, Kim D and Ihm Y 2002 *Appl. Phys. Lett.* **80** 3358

[89] Han S J, Song J W, Yang C H, Park S H, Park J H, Jeong Y H and Rhie K W 2002 *Appl. Phys. Lett.* **81** 4212

[90] Lee H J, Jeong S Y, Hwang J Y and Cho C R 2003 *Europhys. Lett.* **64** 797

[91] Fukumura T, Yamada Y, Tamura K, Nakajima K, Aoyama T, Tsukazaki A, Sumiya M, Fuke S, Segawa Y, Chikyow T, Hasegawa T, Koinuma H and Kawasaki M 2003 *Jpn. J. Appl. Phys.* **42** L105

[92]  Toyosaki H, Fukumura T, Yamada Y, Nakajima K, Chikyow T, Hasegawa T, Koinuma H and Kawasaki M cond-mat/0307760

[93]  Higgins J S, Shinde S R, Ogale S B, Venkatesan T and Greene R L 2004 *Phys. Rev.* B **69** 073201

[94]  Shinde S R, Ogale S B, Higgins J S, Zheng H, Millis A J, Kulkarni V N, Ramesh R, Greene R L and Venkatesan T 2004 *Phys. Rev. Lett.* **92** 166601

[95]  Toyosaki H, Fukumura T, Yamada Y, Nakajima K, Chikyow T, Hasegawa T, Koinuma H and Kawasaki M 2004 *Nat. Mater.* **3** 221

[96]  Stampe P A, Kennedy R J, Xin Y and Parker J S 2003 *J. Appl. Phys.* **93** 7864

[97]  Kim D H, Yang J S, Lee K W, Bu S D, Noh T W, Oh S J, Kim Y W, Chung J S, Tanaka H, Lee H Y and Kawai T 2002 *Appl. Phys. Lett.* **81** 2421

[98]  Seong N J, Yoon S G and Cho C R 2002 *Appl. Phys. Lett.* **81** 4209

[99]  Stampe P A, Kennedy R J, Xin Y and Parker J S 2002 *J. Appl. Phys.* **92** 7114

[100]  Shinde S R, Ogale S B, Das Sarma S, Simpson J R, Drew H D, Lofland S E, Lanci C, Buban J P, Browning N D, Kulkarni V N, Higgins J, Sharma R P, Greene R L and Venkatesan T 2003 *Phys. Rev.* B **67** 115211

[101]  Kim D H, Yang J S, Lee K W, Bu S D, Kim D W, Noh T W, Oh S J, Kim Y W, Chung J S, Tanaka H, Lee H Y, Kawai T, Won J Y, Park S H and Lee J C 2003 *J. Appl. Phys.* **93** 6125

[102]  Yang H S, Choi J Y, Craciun V and Singh R K 2003 *J. Appl. Phys.* **93** 7873

[103]  Li J, Sow O H, Rao S X, Ong C K and Zheng D N 2003 *Euro. Phys. J.* B **32** 471

[104]  Manivannan A, Seehra M S, Majumder S B and Katiyar R S 2003 *Appl. Phys. Lett.* **83** 111

[105]  Manivannan A, Glaspell G and Seehra M S 2003 *J. Appl. Phys.* **94** 6994

[106] Yang H S, Choi J, Song S J and Singh R K 2004 *Electrochem. Solid State Lett.* **7** C4

[107] Kim D H, Yang J S, Kim Y S, Chang Y J, Noh T W, Bu S D, Kim Y W, Park Y D, Pearton S J and Park J H 2004 *Annal. Phys.* **13** 70

[108] Balagurov L A, Klimonsky S O, Kobeleva S P, Orlov A F, Perov N S and Yarkin D G 2004, *JETP Lett.* **79**, 98

[109] Seong N J and Yoon, S G 2004 *J. Electrochem. Soc.* **151** G227

[110] Yang H S and Singh R K 2004 *J. Appl. Phys.* **95** 7192

[111] Nguyen H H, Prellier W, Sakai J and Ruyter A 2004 *J. Appl. Phys.* **95** 7378

[112] Yamada Y, Toyosaki H, Tsukazaki A, Fukumura T, Tamura K, Segawa Y, Nakajima K, Aoyama T, Chikyow T, Hasegawa T, Koinuma H and Kawasaki M 2004 *J. Appl. Phys.* **96** 5097

[113] Murakami M, Matsumoto Y, Nagano M, Hasegawa T, Kawasaki M and Koinuma H 2004 *Appl. Surf. Sci.* **223** 245

[114] Park W K, Ortega-Hertogs R J, Moodera J S, Punnoose A and Seehra M S 2002 *J. Appl. Phys.* **91** 8093

[115] Rameev B Z, Yildiz F, Tagirov L R, Aktas B, Park W K and Moodera J S 2003 *J. Mag. Mag. Mat.* **258** 361

[116] Punnoose A, Seehra M S, Park W K and Moodera J S 2003 *J. Appl. Phys.* **93** 7867

[117] Joh Y G, Kim H D, Kim B Y, Woo S I, Moon S H, Cho J H, Kim E C, Kim D H and Cho C R 2004 *J. Korea. Phys. Soc.* **44** 360

[118] Kennedy R J, Stampe P A, Hu E H, Xiong P, von Molnar S and Xin Y 2004 *Appl. Phys. Lett.* **84** 2832

[119] Shinde S R, Ogale S B, Higgins J S, Zheng H, Millis A J, Kulkarni V N, Ramesh R, Greene R L and Venkatesan T 2004 *Phys. Rev. Lett.* **92** 166601

[120] Seong N J and Yoon S G 2004 *J. Vac. Sci. Tech.* B **22** 762

[121] Chambers S A, Thevuthasan S, Farrow R F C, Marks R F, Thiele J U, Folks L, Samant M G, Kellock A J, Ruzycki N, Ederer D L and Diebold U 2001 *Appl. Phys. Lett.* **79** 3467

[122] Soo Y L, Kioseoglou G, Kim S, Kao Y H, Devi P S, Parise J, Gambino R J and Gouma P I 2002 *Appl. Phys. Lett.* **81** 655

[123] Chambers S A, Wang C M, Thevuthasan S, Droubay T, McCready D E, Lea A S, Shutthanandan V and Windisch C F 2002 *Thin Solid Films* **418** 197

[124] Kim J Y, Park J H, Park B G, Noh H J, Oh S J, Yang J S, Kim D H, Bu S D, Noh T W, Lin H J, Hsieh H H and Chen C T 2003 *Phys. Rev. Lett.* **90** 17401

[125] Shimizu N, Sasaki S, Hanashima T, Yamawaki K, Udagawa H, Kawamura N, Suzuki M, Maruyama H, Murakami M, Matsumoto Y and Koinuma H 2004 *J. Phys. Soc. Jpn.* **73** 800

[126] Lussier A, Dvorak J, Idzerda Y U, Ogale S B, Shinde S R, Choudary R J and Venkatesan T 2004 *J. Appl. Phys.* **95** 7190

[127] Murakami M, Matsumoto Y, Hasegawa T, Ahmet P, Nakajima K, Chikyow T, Ofuchi H, Nakai I and Koinuma H 2004 *J. Appl. Phys.* **95** 5330

[128] Wang Z J, Tang J K, Zhang H G, Golub V, Spinu L and Tung L 2004 *J. Appl. Phys.* **95** 7381

[129] Hong N Y H, Sakai J and Hassini A 2004 *Appl. Phys. Lett.* **84** 2602

[130] Wang W, Dai J, Tang J, Jiang D T, Chen Y, Fang J, He J, Zhou W and Spinu L 2003 *J. Supercond.* **16** 155

[131] Wang Z J, Tang J K, Tung L D, Zhou W L, Spinu L 2003 *J. Appl. Phys.* **93** 7870

[132] Wang Z J, Wang W D, Tang J K, Tung L D, Spinu L and Zhou WL 2003 *Appl. Phys. Lett.* **83** 518

[133] Kim Y J, Thevuthasan S, Droubay T, Lea A S, Wang C M, Shutthanandan V, Chambers S A, Sears R P, Taylor B and Sinkovic B 2004 *Appl. Phys. Lett.* **84** 3531

[134] Lee H M, Kim S J, Shim I B and Kim C S 2003 *IEEE Trans. Magn.* **39** 2788

[135] Hong N H, Prellier W, Sakai J and Hassini A 2004 *Appl. Phys. Lett.* **84** 285

[136] Kimura H, Fukumura T, Koinuma H and Kawasaki M 2001 *Physica* E **10** 265

[137] Kimura H, Fukumura T, Kawasaki M, Inaba K, Hasegawa T and Koinuma H 2002 *Appl. Phys. Lett.* **80** 94

[138] Park Y R and Kim K J 2003 *J. Appl. Phys.* **94** 6401

[139] Fitzgerald C B, Venkatesan M, Douvalis A P, Huber S, Coey J M D and Bakas T 2004 *J. Appl. Phys.* **95** 7390

[140] Coey J M D, Douvalis A P, Fitzgerald C B and Venkatesan M 2004 *Appl. Phys. Lett.* **84** 1332

[141] Ogale S B, Choudhary R J, Buban J P, Lofland S E, Shinde S R, Kale S N, Kulkarni V N, Higgins J, Lanci C, Simpson J R, Browning N D, Das Sarma S, Drew H D, Greene R L and Venkatesan T 2003 *Phys. Rev. Lett.* **91** 077205

[142] Shim I B and Kim C S 2004 *J. Magn. Magn. Mat.* **272-276** 1571

[143] Philip J, Theodoropoulou N, Berera G, Moodera J S and Satpati B 2004 *Appl. Phys. Lett.* **85** 777

[144] Ando K, Takahashi K, Okuda T and Umehara M 1992 *Phys. Rev.* B **46** 12289

[145] In Fig. 2a of Ref [19], the measurement temperatures for the absorption and

MCD spectra are different. Consequently, the absorption spectrum around the absorption edge is very broad in comparison with the MCD spectrum.

[146] Ando K cond-mat/0208010

[147] *The Hall Effect and its Applications* ed Chien C L and Westgate C R (New York: Plenum 1980)

[148] Aronzon B A, Granovskii A B, Kovalev D Y, Meilikhov E Z, Ryl'kov V V and Sedova M V 2000 *JETP Lett.* **71** 687

[149] Pakhomov A B, Yan X and Xu Y 1996 *J. Appl. Phys.* **79** 6140

[150] Toyosaki H, Yamada Y, Fukumura T and Kawasaki M unpublished.

Table 1   Properties of typical ferromagnetic conductors.

| Material | $T_C$ | Polarization | Carrier density | |
|---|---|---|---|---|
| Ferromagnetic metal | ~$10^3$ K | 10-30 % | $10^{23}$ cm$^{-3}$ | |
| Ferromagnetic semiconductor | ~200 K | ~80 % | $10^{16}$-$10^{22}$ cm$^{-3}$ | ($p$-type) |
| Ferromagnetic perovskite oxide | < 400 K | ~100 % | $10^{22}$ cm$^{-3}$ | ($p$-type) |
| Ferromagnetic oxide semiconductor | > 300 K? | ? | $10^{18}$-$10^{22}$ cm$^{-3}$ | ($n$-type) |

Table 2  Electronic properties of various transparent oxide semiconductors ([10,95,112]), where $\rho$, $n$, $\mu$, and $E_g$ denote resistivity, carrier density, mobility, and energy gap at room temperature, respectively.

| Crystal structure | Material | $\rho$ [Ohm.cm] | $n$ [cm$^{-3}$] | $\mu$ [cm$^2$/Vs] | Polarity | $E_g$ [eV] |
|---|---|---|---|---|---|---|
| Rutile | $SnO_2$ | $7.5\times10^{-5}$-$7.5\times10^{-4}$ | $2.7\times10^{20}$-$1.2\times10^{21}$ | 18-31 | $n$ | 3.8-4.0 |
| | $TiO_2$ | $3\times10^{-3}$-$2\times10^{1}$ | $10^{18}$-$10^{21}$ | 0.05-0.2 | $n$ | 3.0-3.5 |
| Anatase | $TiO_2$ | $6\times10^{-2}$-$8\times10^{-2}$ | $7\times10^{18}$-$2\times10^{19}$ | 6-10 | $n$ | 3.0-3.5 |
| NaCl | CdO | $1.30\times10^{-4}$ | $1.00\times10^{21}$ | 48 | $n$ | 2.2-3.2 |
| Spinel | $CdIn_2O_4$ | $2.70\times10^{-4}$ | $4.00\times10^{20}$ | 57 | $n$ | 3.24 |
| | $MgIn_2O_4$ | $4.3\times10^{-3}$-$2.0\times10^{-1}$ | $2.9\times10^{19}$-$6.3\times10^{20}$ | 1.1-3.0 | $n$ | 3.24 |
| C-rare earth | $In_2O_3$ | $4.30\times10^{-5}$ | $1.40\times10^{21}$ | 103 | $n$ | 3.5-4.0 |
| Wurtzite | ZnO | $1.9\times10^{-4}$-$5.1\times10^{-4}$ | $1.1\times10^{20}$-$1.5\times10^{21}$ | 28-120 | $n$ | 3.3-3.6 |
| $\beta$-$Ga_2O_3$ | $Ga_2O_3$ | $2.60\times10^{-2}$ | $5.20\times10^{18}$ | 46 | $n$ | 4.8-5.0 |
| $Sr_2PbO_4$ | $Cd_2SnO_4$ | $1.30\times10^{-4}$ | $8.00\times10^{20}$ | 65 | $n$ | 2.1-2.9 |
| $YbFe_2O_4$ | $InGaMgO_4$ | 1.89 | $1.75\times10^{18}$ | 1.9 | $n$ | 4.2 |
| | $InGaZnO_4$ | $7.25\times10^{-3}$ | $5.21\times10^{19}$ | 16.5 | $n$ | 3.5 |
| Pyrochlore | $AgSbO_3$ | 3.4 | $2.80\times10^{18}$ | 7.5 | $n$ | 3.1 |
| | $Cd_2Sb_2O_7$ | $2.40\times10^{-2}$ | $1.30\times10^{20}$ | 1.9 | $n$ | 3.5 |
| Olivine | $Cd_2GeO_4$ | $6.7\times10^{-3}$-$3.3\times10^{-1}$ | $1.3\times10^{18}$-$1.0\times10^{20}$ | 9.2 | $n$ | 3.4 |
| Delafossite | $AgInO_2$ | $1.70\times10^{-1}$ | $2.70\times10^{19}$ | 0.47 | $n$ | 3.4 |
| | $CuAlO_2$ | 11 | $1.30\times10^{17}$ | 10.4 | $p$ | 3.5 |
| | $CuGaO_2$ | $1.80\times10^{2}$ | $2.40\times10^{18}$ | $1.50\times10^{-2}$ | $p$ | 3.4 |
| $SrCu_2O_2$ | $SrCu_2O_2$ | 21 | $6.10\times10^{17}$ | 0.46 | $p$ | 3.3 |

Table 3　Advantages of oxide semiconductors.

| Merits |
| --- |
| Wide gap |
| 　　Transparent |
| 　　Short wavelength applications |
| 　　Colorable |
| Can grow on plastic |
| Ecologically safe |
| Low cost fabrication (high Clarke number) |
| Durable |
| High carrier density ($n$-type) |

Table 4  Recent papers that report various ferromagnetic semiconductors based on ZnO. Right column denotes the number of the papers of ferromagnetic ZnO based semiconductors in this table and the total papers including nonferromagnetic ZnO based semiconductors.

| Year | $T_C$ of ferromagnetic transition metal-doped ZnO | | | | | | | | Number of papers (total) |
|---|---|---|---|---|---|---|---|---|---|
| | Ti | V | Cr | Mn | Fe | Co | Ni | Cu | |
| 1999 | - | - | - | - | - | - | - | - | 0 (1) |
| 2000 | - | - | - | - | - | - | - | - | 0 (1) |
| 2001 | - | >300-350K [29,30] | - | - | - | >300K [30,32] | 2 K [83] | - | 4 (10) |
| 2002 | - | - | - | 45K [33] | >300K with Co [88]<br>550K with Cu [89] | >350K [62,64]<br>>300K with Fe [88] | - | with Fe [89] | 5 (10) |
| 2003 | - | - | - | >r.t. [41]<br>37K [42]<br>70K [44]<br>>250K [52]<br>225-300K [53]<br>250K with Sn [87] | - | 225-300K [53]<br>>350K [56]<br>>r.t. [66,69]<br>>350K with Li [90] | - | - | 10 (17) |
| 2004 | - | - | - | 50K [47]<br>r.t. [48] | - | >r.t. [78]<br>150K-r.t. [82] | - | - | 4 (15)* |

\* As of summer.

Table 5  Recent papers that claim extrinsic origin of ferromagnetism for Co-doped TiO$_2$. Right column denotes the numbers of the papers in the center column and the total papers in each year.

| Year | Extrinsically ferromagnetic Co-doped TiO$_2$ | | Number of papers (total) |
|---|---|---|---|
| | anatase | rutile | |
| 2001 | - | - | 0 (3) |
| 2002 | Refs. [97,99] | - | 2 (7) |
| 2003 | Refs. [96,101,104,124] | Refs. [115,116] | 6 (16) |
| 2004 | Refs. [107,110] | Refs. [94,118,133] | 5 (18)* |

* As of summer.

**Figure captions**

Figure 1. Schematic digest of the $A_{1-x}Mn_xB$ (A: II group element; B: VI group element) magnetic semiconductors and their crystal structures (cf. Ref. [1]). The bold lines represent the range of $x$ of their single phases. Cub and hex denote cubic and hexagonal crystal structures, respectively. Mn-doped CdO and Mn-doped HgO have not been reported.

Figure 2. Schematic diagrams of intrinsic and extrinsic ferromagnetic origins for (a) magnetization, (b) magneto-optical effect, and (c) magneto-transport measurements. Left and center columns represent the intrinsic and extrinsic ferromagnetism, respectively. Right column represent the phenomenological scheme for each measurement.

Figure 3. (a) Magnetic circular dichroism (MCD) and absorption spectra for Co-doped ZnO (dashed line) and Co-doped anatase $TiO_2$ (solid line). Measurement temperature is 5 K for Co-doped ZnO and 300 K for Co-doped $TiO_2$. Co content is 2 mol% for MCD spectrum and 1 mol% for absorption spectrum of Co-doped ZnO, and 10 mol% for Co-doped $TiO_2$. (b) Magnetic field dependence of MCD for Co-doped ZnO at 6 K with incident photon energy of 1.94 eV and for Co-doped $TiO_2$ at 300 K with incident photon energy of 3.57 eV. Co content is 10 mol% for Co-doped ZnO and $TiO_2$.

Figure 4. Magnetic field dependence of Hall resistivity for rutile Co-doped $TiO_2$ at 300 K. The dashed line represents ordinary part of the Hall resistivity, corresponding to the electron density of $4 \times 10^{21}$ cm$^{-3}$. Upper inset is a photograph of Hall bar for the measurement, and bottom inset is magnetic field dependence of the magnetization for the same sample at 300 K.

Figure 5. Mapping for the appearance of ferromagnetic Co-doped rutile TiO$_2$ as functions of the conductivity and Co content determined from the anomalous Hall effect and MCD measurements at 300 K. Circle, triangle, and cross symbols respectively represent the appearance of anomalous Hall effect and ferromagnetic MCD, weak anomalous Hall effect and ferromagnetic MCD, and disappearance of anomalous Hall effect and MCD.

Figure 6. Relationship between the anomalous part of Hall conductivity ($\sigma_{AH}$) and conductivity ($\sigma_{xx}$) for Co-doped rutile TiO$_2$. These plots include the data for different Co contents, measurement temperatures, and growth oxygen pressures $P_{O2}$ (*i.e.* carrier density). Rectangle, circle, and triangle plots represent the data for $P_{O2} = 10^{-8}$ Torr, $10^{-7}$ Torr, and $10^{-6}$ Torr, respectively. Solid symbols with solid line, open symbols with solid line, and solid symbols with dashed line correspond to the data for $x = 0.03$, 0.05, and 0.10, respectively. The data at 100 K, 200 K, and 300 K from the same sample are connected with lines. The data by Higgins et al. [93] are also plotted.

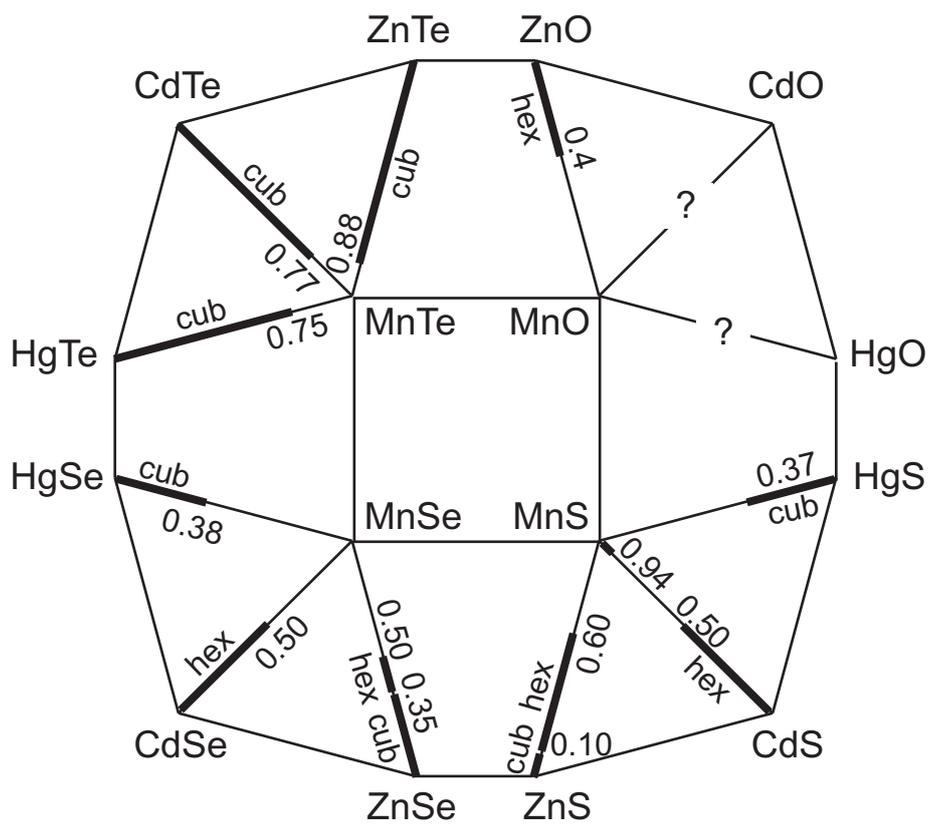

Figure 1  T. Fukumura

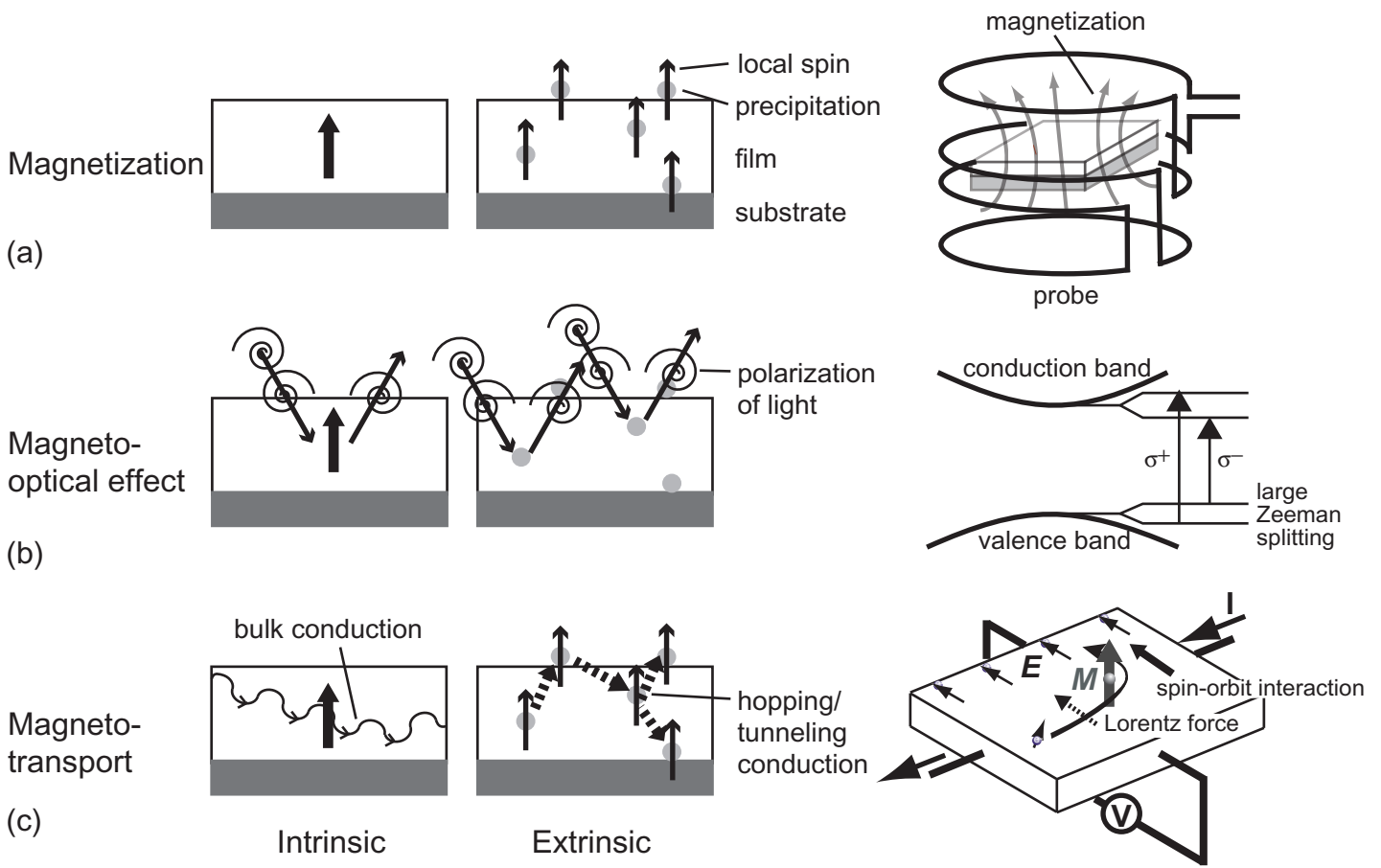

Figure 2 T. Fukumura

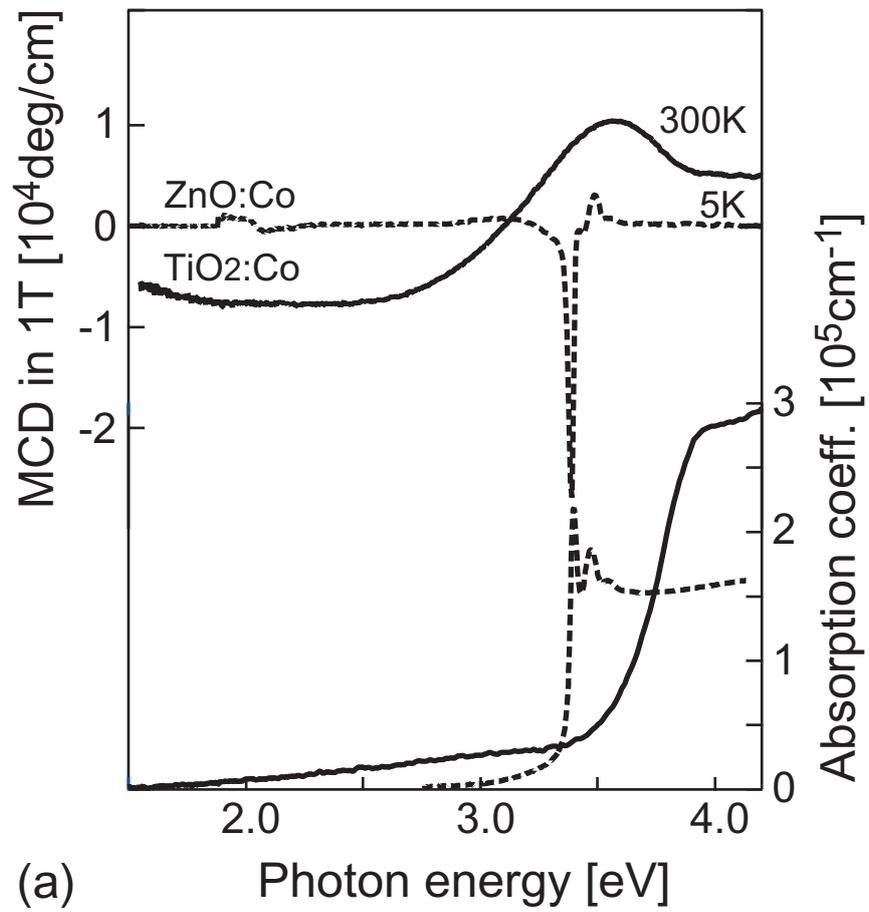

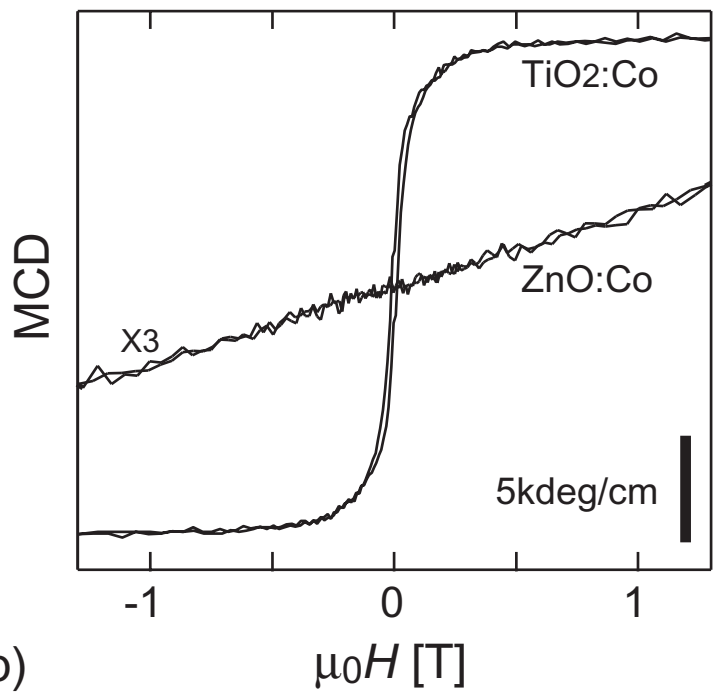

Figure 3  T. Fukumura

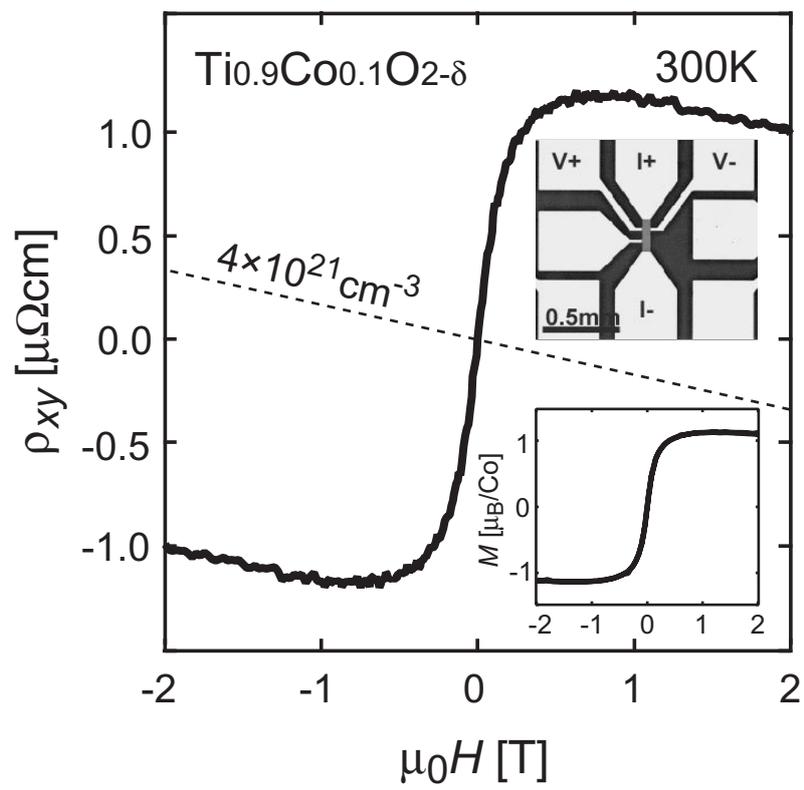

Figure 4  T. Fukumura

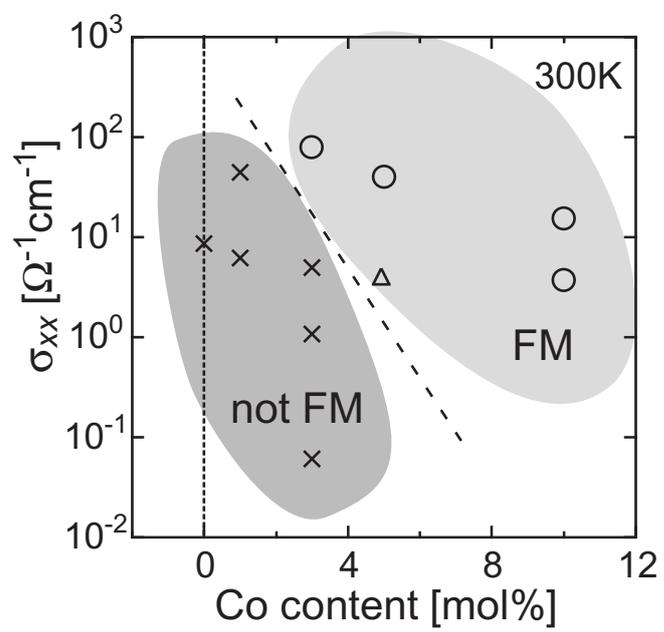

Figure 5 T. Fukumura

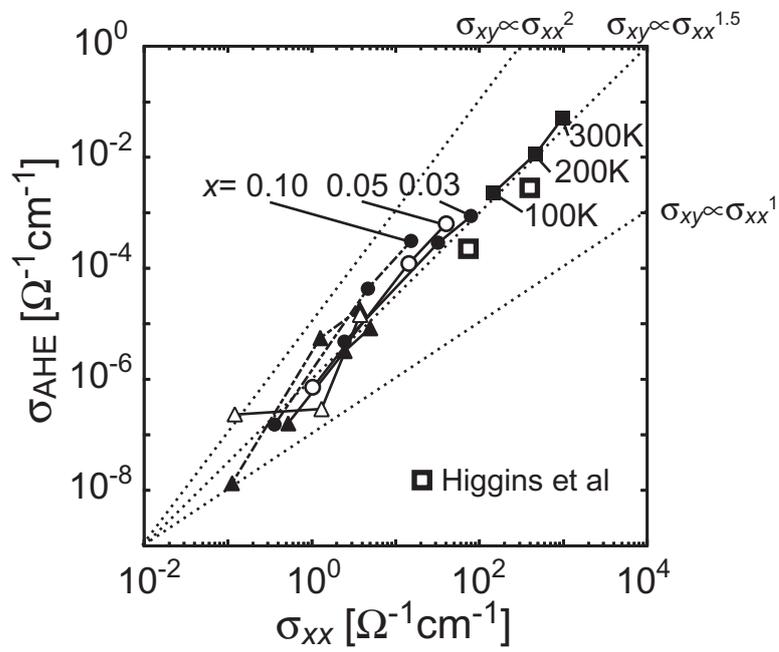

Figure 6  T. Fukumura